\documentclass[smallextended,pdflatex,sn-mathphys-num]{sn-jnl}

\usepackage{graphicx}%
\usepackage{multirow}%
\usepackage{amsmath,amssymb,amsfonts}%
\usepackage{amsthm}%
\usepackage{mathrsfs}%
\usepackage[title]{appendix}%
\usepackage{xcolor}%
\usepackage{textcomp}%
\usepackage{manyfoot}%
\usepackage{booktabs}%
\usepackage{algorithm}%
\usepackage{algorithmicx}%
\usepackage{algpseudocode}%
\usepackage{listings}%
\usepackage{array}%

\usepackage{subcaption}
\usepackage{hyperref}

\theoremstyle{thmstyleone}%
\theoremstyle{thmstyletwo}%

\theoremstyle{thmstylethree}%

\begin{document}

\title[Data Provenance Aggregation and Structural Descriptors for Query-Template Identification]{Data Provenance Aggregation and Structural Descriptors for Query-Template Identification}


\author*[1]{\fnm{Paulo} \sur{Pintor}}
\email{paulopintor@ua.pt}

\author[2]{\fnm{Rogério Luís} \sur{Costa}}
\email{rogcosta@dei.uc.pt}

\author[1]{\fnm{José} \sur{Moreira}}
\email{jose.moreira@ua.pt}

\affil[1]{%
  \orgdiv{IEETA},
  \orgname{University of Aveiro},
  \orgaddress{\city{Aveiro}, \country{Portugal}}
}

\affil[2]{%
  \orgdiv{CISUC, DEI},
  \orgname{University of Coimbra},
  \orgaddress{\city{Coimbra}, \country{Portugal}}
}


\abstract{ 
Runtime query similarity identification supports several activities, including workload analysis, identifying recurring behaviour, auditing, debugging, and anomaly detection. Existing methods for evaluating query similarity mostly rely on SQL text, execution plans, or result values but do not account for input data or how results are derived during query execution.

This paper proposes a novel approach using data provenance for query-template identification. We construct query-level profiles that capture explicit data dependencies, occurrence frequencies, and structural properties by aggregating tuple-level Lineage and Why-provenance across entire query results. We also derive Provenance Structural Descriptors (PSD), which summarise these profiles using a small set of numerical features.

We evaluate our proposals using these profiles to identify query templates over \(12{,}500\) distinct query instances generated from parametrised TPC-DS query templates. The experimental evaluation shows that query templates can be identified solely from their provenance footprints, and that accuracy improves as more historical data becomes available. It also shows that PSD achieve high classification accuracy with limited training history and low test-processing cost.
}

\keywords{Data provenance, Query similarity, Query-level provenance, Lineage, Why-provenance} 



\maketitle

\section{Introduction}
\label{sec:intro}

Modern data systems execute large numbers of queries, many of which are generated from a smaller set of parametrised templates representing routine tasks or user intents~\cite{10.14778/3430915.3430931,Moreau2022,10.14778/3750601.3750632}. 
Given this templated structure, characterising queries and measuring their similarity is important to  cluster and summarise workloads~\cite{10.14778/3430915.3430931,10.14778/3750601.3750632}, identify recurring execution patterns~\cite{lai_edbt_2023,Moreau2022}, or relate new queries to historical workloads~\cite{frost_edbt_2024}, as well as for downstream applications such as auditing, debugging, user profiling, and anomaly detection~\cite{10.1145/3514221.3517861,bertino2019}. 


Existing approaches to characterising queries and measuring their similarity obtain their features from the query text or its abstract syntax tree~\cite{Makiyama2015TextMA, Kul2018}, from execution plans and runtime statistics~\cite{Paul2021}, or from the results the queries return~\cite{8700244, ARZAMASOVA2021101646}, including the the most recent AI-based methods~\cite{Paul2021, tang2022}. While effective at capturing how a query is written, executed, or what it returns, these features provide limited insight into how its results derive from the underlying data: queries with different text may produce the same results through different derivations, whereas queries with similar text or results may access and combine data in very different ways.

Data provenance formally describes the relationship between a query's output and the underlying source data. Rather than recording only the result, provenance models capture information about how each output tuple is derived. Depending on the model, this information may identify contributing input tuples, explain why an output tuple appears, or describe how alternative derivations combine~\cite{Buneman2001,Green2007,Cheney2007,Herschel2017,Pierre2018}. This offers a way to characterise queries based on their underlying data dependencies, complementing the features used by existing approaches and providing a basis for measuring query similarity and classifying queries according to how their results are derived.

In this work, we investigate query-template identification using data provenance.
The task is to predict the parametrised query template that generated a given query instance.
A template is a parametrized query where constants are replaced by variables.
Our hypothesis is that there are template-specific characteristics that remain consistent across parameter instantiations and discriminative across different templates. To make it usable for query-template identification, we aggregate tuple-level provenance across each query result into query-level representations, without requiring correspondences between individual output tuples. The resulting representations make contributing tuples and derivational structure available as features for query-template identification, allowing the task to be addressed through observed data dependencies rather than SQL text, execution plans, or result values.

We consider two established provenance abstractions: Lineage, which records the input tuples contributing to an output tuple, and Why-provenance, which additionally captures alternative derivations and their conjunctive structure through witnesses. We formalize set-based aggregations, which record the presence of contributing tuples or witnesses, and bag-based aggregations, which capture their occurrence frequencies across the query result. Building on these abstractions, we investigate two complementary approaches to comparing query executions.
The first examines structural representations that preserve varying degrees of derivational detail, ranging from flat contributing tuples to atomic or blocked witness compositions, using specialized similarity measures.
The second encodes provenance characteristics as numerical feature vectors, including both fine-grained identity-based representations and a novel set of compact, fixed-dimension metrics termed Provenance Structural Descriptors (PSD), enabling query-template classification via standard nearest-neighbor and supervised machine learning methods.
The main contributions are:

\begin{itemize}
    \item We formalise set-based and bag-based aggregations of tuple-level Lineage and Why-provenance for characterising and classifying complete query executions.
    \item We define and evaluate provenance representations that preserve different information, including contributing tuple identities, complete witness identities, occurrence frequencies, and compact structural descriptors.
    \item We introduce a candidate-restricted comparison procedure for Why-provenance that recognises partial witness overlap while limiting the query- and witness-level comparisons performed.
    \item We analyse classification effectiveness and test-processing cost across classifiers and training-history sizes, and assess PSD feature contributions through model attributions, descriptor-group comparisons, and single-descriptor experiments.
\end{itemize}

The results show that query-level provenance provides an effective basis for query-template identification, offering a complementary perspective based on how database tuples contribute to the derivation of query results.  This perspective makes explicit relationships that are not disclosed in previous work, such as queries producing identical results from different contributing tuples. The proposed framework can be deployed independently to classify queries according to their actual data usage or integrated with syntax-, plan-, and result-based techniques. This integration enables workload characterization from multiple perspectives and can improve the identification of recurring query patterns for anomaly detection, user profiling and other analysis tasks.


The remainder of this paper is organised as follows. Section~\ref{sec:background-related-work} introduces the relevant provenance concepts, reviews related work, and presents motivating examples. Section~\ref{sec:representations} formalises the query-level aggregations and defines the resulting representations and comparison procedures. Section~\ref{sec:results} describes the experimental setup and implementation, presents the results, and discusses their implications. Finally, Section~\ref{sec:conclusion} summarises the contributions and outlines directions for future work.
\section{Related Work}
\label{sec:background-related-work}

This section introduces the data provenance concepts underlying the representations considered in this work and reviews existing approaches to query similarity and workload analysis. It first presents the provenance models, then examines how previous studies have compared and characterised queries, and finally illustrates the complementary information  provenance provides when comparing query executions.

\subsection{Data Provenance}
\label{subsec:data-provenance}

Data-provenance models formalise how query results are derived from the input data, capturing different aspects of their origin and derivation~\cite{Cheney2007,Herschel2017,Buneman2018,Glavic2021}. Within the semiring framework, provenance polynomials provide an algebraic representation of derivations for positive relational queries~\cite{Green2007,Green2017}. Other provenance models can be obtained as abstractions of these polynomials, retaining different aspects of the contributing data and their participation in derivations~\cite{Green2009}.

In this work, we focus on Lineage, which records the input tuples contributing to an output tuple, and Why-provenance, which additionally preserves their grouping into alternative derivations. These models distinguish between identifying the contributing tuples and recording how those tuples participate jointly in producing a result.

Let \(Q\) be a positive relational query, \(I\) a database instance, and \(Q(I)\) the result of evaluating \(Q\) over \(I\). Each input tuple in \(I\) is identified by a unique provenance token. We denote an output tuple by \(t \in Q(I)\).

\medskip
\noindent\textbf{Why-provenance.} This representation records the combinations of input tuples that participate in deriving an output tuple~\cite{Buneman2001,Green2009}. A witness is the set of provenance tokens identifying the input tuples used jointly in one derivation. Let \(\mathcal{D}(Q,I,t)\) denote the derivations of \(t\), and let \(\operatorname{supp}(d)\) denote the set of provenance tokens used in derivation \(d\). The Why-provenance of \(t\) is defined as
\[
\mathrm{Wit}(Q,I,t)
=
\left\{
\operatorname{supp}(d)
\mid
d \in \mathcal{D}(Q,I,t)
\right\}.
\]
Each witness records which input tuples participate together, without retaining repeated occurrences of a token within the same derivation. Derivations involving the same set of tokens are represented by the same witness.

\medskip
\noindent\textbf{Lineage.} This provenance representation records the input tuples that participate in at least one derivation of an output tuple, without distinguishing how they are grouped into witnesses~\cite{Cheney2007,Green2009}. It is obtained by taking the union of the witnesses:
\[
\mathrm{Lin}(Q,I,t)
=
\bigcup_{w \in \mathrm{Wit}(Q,I,t)} w.
\]
Both representations are set-valued at the tuple level. Why-provenance retains distinct witnesses, whereas Lineage combines their contributing tokens into a single set.

Figure~\ref{fig:running-example-multiwitness} shows the database instance used as a running example throughout this work. The \textit{Customers} and \textit{Orders} relations each include a column containing the unique provenance token assigned to every input tuple.

\begin{figure}[htb]
\centering\small
\begin{minipage}[t]{0.48\linewidth}
\centering\textbf{\textit{Customers}}\par\medskip
\begin{tabular}{ccc}
\toprule
\textbf{Cust} & \textbf{Segment} & \textbf{Token} \\
\midrule
\(c_1\) & Retail & \(t_1\) \\
\(c_2\) & Retail & \(t_2\) \\
\bottomrule
\end{tabular}
\end{minipage}\hfill%
\begin{minipage}[t]{0.48\linewidth}
\centering\textbf{\textit{Orders}}\par\medskip
\begin{tabular}{ccc}
\toprule
\textbf{Cust} & \textbf{Status} & \textbf{Token} \\
\midrule
\(c_1\) & Open & \(t_3\) \\
\(c_2\) & Open & \(t_4\) \\
\(c_1\) & Shipped & \(t_5\) \\
\bottomrule
\end{tabular}
\end{minipage}
\caption{Database instance for the running example.}
\label{fig:running-example-multiwitness}
\end{figure}

Consider query \(Q_1\), which joins the two relations on the customer identifier and returns the distinct order statuses: 

\medskip
\noindent\textbf{Query \(Q_1\).}
\begin{quote}
\small
\begin{verbatim}
SELECT DISTINCT o.Status
FROM Customers AS c
JOIN Orders AS o ON c.Cust = o.Cust;
\end{verbatim}
\end{quote}

Figure~\ref{fig:running-example-result} presents the result together with its Why-provenance and Lineage. The output tuple \(\mathit{Status}=\texttt{'Open'}\) has two alternative derivations. One joins the customer tuple identified by \(t_1\) with the order tuple identified by \(t_3\), forming the witness \(\{t_1,t_3\}\). The other joins the tuples identified by \(t_2\) and \(t_4\), forming the witness \(\{t_2,t_4\}\). Its Lineage is therefore \(\{t_1,t_2,t_3,t_4\}\). The output tuple \(\mathit{Status}=\texttt{'Shipped'}\) has a single derivation, joining the tuples identified by \(t_1\) and \(t_5\). Its Why-provenance is \(\{\{t_1,t_5\}\}\), and its Lineage is \(\{t_1,t_5\}\).

\begin{figure}[htb]
\centering\small
\begin{tabular}{ccc}
\toprule
\textbf{Status} & \textit{Why} & \textit{Lineage} \\
\midrule
Open & \(\{\{t_1,t_3\},\{t_2,t_4\}\}\) & \(\{t_1,t_2,t_3,t_4\}\) \\
Shipped & \(\{\{t_1,t_5\}\}\) & \(\{t_1,t_5\}\) \\
\bottomrule
\end{tabular}
\caption{Result of \(Q_1\) with Why-provenance and Lineage.}
\label{fig:running-example-result}
\end{figure}

\subsection{Query Similarity and Workload Analysis}
\label{subsec:query-similarity-related}

Previous work on query similarity has primarily focused on analysing SQL logs to group statements with related structures or intents.
These approaches commonly represent queries using structural features extracted from selections, joins, projections, and aggregations, and compare them using measures such as Hamming distance, Jaccard similarity, or cosine similarity~\cite{Aligon2014,Makiyama2015TextMA}. For example, the approach proposed in~\cite{Kul2018} clusters queries using structural heuristics and query rewriting, without requiring access to the database schema or underlying data.
Moreau et al.~\cite{Moreau2022} use a similar approach to cluster queries and identify recurring patterns of user behaviour on a database-as-a-service platform.
%
Other methods, like Deep et al.~\cite{10.14778/3430915.3430931}, represent SQL queries through categorical features extracted from the query text and numerical execution statistics, and select a reduced workload that preserves coverage and representativeness. Their objective is to retain the global characteristics of a workload rather than compare individual query executions according to the data dependencies underlying their results.

Another line of research compares SQL queries according to the tuples contained in their results. Result-based similarity commonly uses descriptive statistics or set-overlap measures, such as Jaccard similarity, to quantify the relationship between result sets. Arzamasova et al.~\cite{8700244} examine the effectiveness of these measures for identifying user interests and compare them with structure-based alternatives. Result-set similarity has also been used in query recommendation systems to retrieve queries associated with related information needs~\cite{ARZAMASOVA2021101646}. In anomaly detection, result-derived access patterns have been used to model historical behaviour and identify unusual aggregations or unauthorised access to relational data~\cite{bertino2019}.

Learning-based approaches derive numerical representations of SQL queries for downstream tasks. Paul et al.~\cite{Paul2021} learn structural and computational representations from query plans, evaluating them via latency prediction and classification. PreQR~\cite{tang2022} instead combines automaton-based query structures with graph neural networks of the database schema, producing embeddings for clustering and cardinality estimation. Related methods also target anomaly detection in database workloads. UDAD~\cite{Zhong2023} represents SQL statements as semantic vectors, combining bidirectional Long Short-Term Memory (LSTM) networks with an autoencoder to identify anomalous operations. Recently, DBPrompt~\cite{Zhong2024} explored prompt learning to detect and analyse anomalous database activity.

%

Provenance has also been explored in tasks related to query understanding and workload analysis.
Deutch and Gilad~\cite{DeutchGilad2019} study the reverse engineering of conjunctive queries from output tuples and provenance examples, showing that a small number of examples may reveal substantial information about the query that produced them. Gilad and Moskovitch~\cite{GiladMoskovitch2020} investigate query inference from simple explanations that are converted into complete or partial Lineage. They show that information about contributing input tuples can support query inference even when complete derivational information is unavailable. Diestelk{\"a}mper and Herschel~\cite{DiestelkaemperHerschel2020} combine provenance from multiple queries to identify frequently used data items and attributes, as well as attributes that are frequently processed together. These patterns support auditing and data-layout optimisation. The analysis therefore characterises how a workload uses the underlying data, rather than classifying the individual query executions.

On the other hand, Deutch et al.~\cite{Deutch2021Privacy} address the privacy concerns raised by the detailed information provenance exposes. They study how to abstract provenance so that it remains useful for explaining results while concealing the underlying query.

Their approach replaces tuple annotations with more general annotations, balancing information loss against the number of queries consistent with the disclosed provenance. This work highlights that provenance can distinguish possible query derivations beyond what is revealed by the output tuples alone.

Arad et al.~\cite{Arad2022} use historical query logs, output tuples, and Shapley-based fact-contribution values to predict contribution rankings for new query results. Their approach identifies relevant historical queries using syntactic similarity, result similarity, and similarity between fact-contribution rankings.
However, Shapley values quantify how much each fact contributes to a result's value, which is a different notion from provenance, which records how the result was derived from the input.

The query-similarity approaches reviewed above characterise queries primarily through their expressions, execution plans, schema features, or observed results. These representations do not explicitly describe the particular input tuples contributing to each result and their grouping into derivations. The provenance-related studies address different questions, including query inference, privacy, fact contributions, and workload-wide data usage. The question investigated in this work is whether tuple-level Lineage and Why-provenance can be aggregated into query-level representations that retain sufficiently discriminative information for query-template identification.

\subsection{Motivation for Provenance-Based Query Comparison}
\label{subsec:provenance-comparison-motivation}

The following examples use the database instance in Figure~\ref{fig:running-example-multiwitness} to illustrate the information that provenance makes available for query comparison and its relevance to query-template identification. They distinguish between the values returned by a query, the input tuples contributing to those values, and the combinations of tuples involved in their derivation. In the proposed approach, classification uses representations constructed from materialised provenance, rather than SQL text, execution plans, schema features, or result values.

\smallskip
\noindent\textbf{Identical results with different data dependencies.} Consider the following queries returning  the same result: $Q_2(I)=Q_3(I)=\{\texttt{Open}\}$. 
\begin{center}
\begin{minipage}[h]{0.48\textwidth}
\textbf{Query \(Q_2\).}
\begin{quote}
\small
\begin{verbatim}
SELECT DISTINCT o.Status
FROM Orders AS o
WHERE o.Cust = 'c1'
  AND o.Status = 'Open';
\end{verbatim}
\end{quote}
\end{minipage}
\noindent
\begin{minipage}[h]{0.48\textwidth}
\textbf{Query \(Q_3\).}
\begin{quote}
\small
\begin{verbatim}
SELECT DISTINCT o.Status
FROM Customers AS c
JOIN Orders AS o ON c.Cust = o.Cust
WHERE c.Cust = 'c2'
  AND o.Status = 'Open';
\end{verbatim}
\end{quote}
\end{minipage}
\end{center}
\medskip
Although the result is the same, their data dependencies differ. In \(Q_2\), the output tuple is derived exclusively from the order tuple identified by \(t_3\). In \(Q_3\), it is derived by joining the customer tuple identified by \(t_2\) with the order tuple identified by \(t_4\). Their Why-provenance values are therefore \(\{\{t_3\}\}\) and \(\{\{t_2,t_4\}\}\), respectively, and their Lineage sets are \(\{t_3\}\) and \(\{t_2,t_4\}\).

A comparison based exclusively on result values cannot distinguish these queries on this database instance. Provenance exposes their different contributing tuples and derivations, providing information that may help distinguish their originating templates. Structural comparison can also distinguish the query expressions, but does not directly identify the particular input tuples contributing to each observed result.

\smallskip
\noindent\textbf{Different results from the same query template.} Now consider two instantiations of the same query template, differing only in the selected customer:
\begin{center}    
\begin{minipage}[h]{0.48\textwidth}
\textbf{Query \(Q_4\).}
\begin{quote}
\small
\begin{verbatim}
SELECT DISTINCT o.Status
FROM Customers AS c
JOIN Orders AS o ON c.Cust = o.Cust
WHERE c.Cust = 'c1';
\end{verbatim}
\end{quote}
\end{minipage}
\noindent
\begin{minipage}[h]{0.48\textwidth}
\textbf{Query \(Q_5\).}
\begin{quote}
\small
\begin{verbatim}
SELECT DISTINCT o.Status
FROM Customers AS c
JOIN Orders AS o ON c.Cust = o.Cust
WHERE c.Cust = 'c2';
\end{verbatim}
\end{quote}
\end{minipage}
\end{center}

Their results are:
\[
Q_4(I)=\{\texttt{Open},\texttt{Shipped}\},
\qquad
Q_5(I)=\{\texttt{Open}\}.
\]

For \(Q_4\), the witnesses are \(\{t_1,t_3\}\) for \texttt{Open} and \(\{t_1,t_5\}\) for \texttt{Shipped}. For \(Q_5\), the only witness is \(\{t_2,t_4\}\), supporting \texttt{Open}. Taking the union of the Lineage sets of the output tuples gives \(\{t_1,t_3,t_5\}\) for \(Q_4\) and \(\{t_2,t_4\}\) for \(Q_5\).

Although result values, contributing tuples, and witness counts differ, both queries instantiate the same template. Each witness still combines a customer and an order tuple. Thus, template identification cannot require identical provenance across instantiations. Instead, it must exploit characteristics that remain informative despite varying provenance. This motivates examining both representations retaining tuple and witness identities, and descriptors summarising structural provenance properties.

\smallskip
\noindent\textbf{Identical results and contributing tuples with different derivations.} Why-provenance can provide a further distinction by preserving how contributing tuples are combined. Consider query \(Q_1\) introduced in the running example, which joins customers and orders with the same customer identifier, and the following query, which instead joins tuples with different customer identifiers:

\medskip
\noindent\textbf{Query \(Q_6\).}
\begin{quote}
\small
\begin{verbatim}
SELECT DISTINCT o.Status
FROM Customers AS c
JOIN Orders AS o ON c.Cust <> o.Cust;
\end{verbatim}
\end{quote}
\smallskip
Both queries produce the same result: $
Q_1(I)=Q_6(I)=\{\texttt{Open},\texttt{Shipped}\}.
$ For \texttt{Open}, both \(Q_1\) and \(Q_6\)  have the same Lineage, \(\{t_1,t_2,t_3,t_4\}\). For \texttt{Shipped}, the Lineage is \(\{t_1,t_5\}\) in \(Q_1\) and \(\{t_2,t_5\}\) in \(Q_6\). Nevertheless, taking the union of the Lineage sets across each complete result gives the same set of contributing tokens, \(\{t_1,t_2,t_3,t_4,t_5\}\).

The witness collections retain a distinction that this union loses. Collecting the witnesses across the result of \(Q_1\) gives: $
\bigl\{\{t_1,t_3\},\{t_2,t_4\},\{t_1,t_5\}\bigr\},
$
whereas the corresponding collection for \(Q_6\) is $
\bigl\{\{t_1,t_4\},\{t_2,t_3\},\{t_2,t_5\}\bigr\}.
$

Thus, even when the result values and the union of the tuple-level Lineage sets coincide, the witness collections reveal different derivations. This example motivates preserving witness composition when constructing a representation of an entire query, rather than retaining only the union of its contributing tokens.

Together, these examples identify the information gaps addressed by the proposed representations. Result values alone can conceal differences in the input data and derivations responsible for an answer, while retaining only the contributing tokens can conceal differences in how those tokens participate together. Provenance makes these distinctions explicit through tuple identity and witness composition. At the same time, parameter variation shows why recognising a template requires more than exact equality between provenance representations. Our proposal uses this information to characterise queries according to their observed data dependencies and investigates whether the resulting representations support effective query-template identification. The following section formalises the aggregation of tuple-level provenance and defines the query-level representations used for this purpose.

\section{Provenance-Based Query Representations}
\label{sec:representations}

This section defines how tuple-level Lineage and Why-provenance are aggregated to represent complete query executions. It then presents the set-based and vector-based representations derived from these aggregations. Depending on their form, the resulting representations can be compared directly using representation-specific similarity measures or used as feature vectors for vector-based nearest-neighbour and supervised classification methods.

\subsection{Query-Level Provenance Aggregation}
\label{subsec:queryprov}



To characterise complete query executions without requiring correspondences between their output tuples, we aggregate tuple-level Lineage and Why-provenance across each query result. We consider two aggregation strategies: set-based aggregation, recording whether each provenance element occurs at least once, and bag-based aggregation, which additionally preserves occurrence frequencies. These aggregations support identity-based representations and yield the distinct and total counts used to derive Provenance Structural Descriptors.


\subsubsection{Set-based query-level provenance.}

The set-based aggregation records whether a tuple or witness occurs in the provenance of at least one output tuple:

\begin{equation}
\mathrm{Why}(Q,I)
=
\bigcup_{t \in Q(I)}
\mathrm{Wit}(Q,I,t)
\end{equation}

\begin{equation}
\mathrm{Lin}(Q,I)
=
\bigcup_{t \in Q(I)}
\mathrm{Lin}(Q,I,t)
\end{equation}

This aggregation removes multiplicity across the query result. Consequently, two executions involving the same distinct witnesses or input tuples have the same set-based representation even when those elements occur a different number of times. Changes in query parameters or selectivity may nevertheless alter the contributing tuples and witnesses and, therefore, the resulting representation. For the running example in Figure~\ref{fig:running-example-multiwitness}, the set-based query-level representations are:

\[
\mathrm{Why}(Q,I)
=
\bigl\{
\{t_1,t_3\},
\{t_2,t_4\},
\{t_1,t_5\}
\bigr\},
\quad
\mathrm{Lin}(Q,I)
=
\{t_1,t_2,t_3,t_4,t_5\}.
\]

\subsubsection{Bag-based query-level provenance.}

Alternatively, query-level provenance can preserve the number of occurrences of each tuple or witness across the query result. In this case, tuple-level provenance is aggregated using multiset union:

\begin{equation}
\mathrm{Why}_{\mathrm{bag}}(Q,I)
=
\biguplus_{t \in Q(I)}
\mathrm{Wit}(Q,I,t),
\qquad
\end{equation}
\begin{equation}
\mathrm{Lin}_{\mathrm{bag}}(Q,I)
=
\biguplus_{t \in Q(I)}
\mathrm{Lin}(Q,I,t),
\end{equation}

\noindent
where \(\biguplus\) denotes multiset union. This aggregation captures how often tuples and witnesses participate across the query result, thereby retaining frequency and fan-out effects. It may therefore be more sensitive than set-based aggregation to data distributions, result cardinality, and query parameters.
For the running example, the bag-based representations are:
\[
\mathrm{Why}_{\mathrm{bag}}(Q,I)
=
[\{t_1,t_3\},\{t_2,t_4\},\{t_1,t_5\}],
\quad
\mathrm{Lin}_{\mathrm{bag}}(Q,I)
=
[t_1,t_1,t_2,t_3,t_4,t_5].
\]

The token \(t_1\) occurs twice in the bag-based Lineage because it contributes to two output tuples. In this particular example, no witness is repeated, and the Why-provenance therefore contains the same elements under set and bag aggregation. Square brackets are used to denote multisets, allowing repeated elements.

\subsection{Set-Based Provenance Representations}
\label{label:set_based_prov}


Set-based provenance retains each element once, regardless of its frequency across the query result. Lineage represents executions as sets of distinct contributing tuples, whereas Why-provenance uses collections of set-valued witnesses for alternative derivations. We consider two approaches to using this witness information: Atomic Why-provenance, a standalone set representation, and Blocked Why-provenance, a specialised pairwise comparison procedure.

\subsubsection{Lineage Representation}

A query execution is represented by its query-level Lineage set, denoted by \(\mathrm{Lin}(Q,I)\) as defined in Section~\ref{subsec:queryprov}. This set contains the distinct input tuples contributing to at least one tuple in the query result and therefore characterises the portion of the database involved in the computation.

Queries that depend on overlapping subsets of input tuples are expected to produce Lineage sets with substantial overlap. We quantify this using Jaccard similarity:

\begin{equation}
\mathrm{sim}_{LN}(Q_1,Q_2)
=
\frac{
|\mathrm{Lin}(Q_1,I)\cap\mathrm{Lin}(Q_2,I)|
}{
|\mathrm{Lin}(Q_1,I)\cup\mathrm{Lin}(Q_2,I)|
}.
\end{equation}

The resulting value ranges from zero to one. A value of one indicates identical Lineage sets, whereas a value of zero indicates that the query
executions share no contributing input tuples. This set-based representation captures which distinct input tuples contribute to the query result, but not how often each tuple occurs across the output. This multiplicity information is retained separately by the Token Term Frequency representation introduced later.

\subsubsection{Why-Provenance Representations}
\label{subsec:why-representations}

At the query level, Why-provenance provides a richer structural representation than Lineage. Rather than representing a query execution
as a flat set of contributing tuples, it represents it as a collection of witnesses, where each witness contains the input tuples participating jointly in one derivation. Consequently, comparing two query executions requires comparing collections of sets.

To capture partial overlap between derivations, each witness of the first query can be compared with its most similar witness in the second query using the Jaccard index. For two query executions \(Q_1\) and \(Q_2\), the resulting directed similarity is defined as

\begin{equation}
\mathrm{sim}^{\rightarrow}_{\mathrm{WHY}}(Q_1,Q_2)
=
\frac{1}{|\mathrm{Why}(Q_1,I)|}
\sum_{w \in \mathrm{Why}(Q_1,I)}
\max_{v \in \mathrm{Why}(Q_2,I)}
\frac{|w \cap v|}{|w \cup v|}.
\end{equation}

Since this measure is directional, the final similarity is obtained by averaging both directions:

\begin{equation}
    \mathrm{sim}_{\mathrm{WHY}}(Q_1,Q_2)
=
\frac{
\mathrm{sim}^{\rightarrow}_{\mathrm{WHY}}(Q_1,Q_2)
+
\mathrm{sim}^{\rightarrow}_{\mathrm{WHY}}(Q_2,Q_1)
}{2}.
\end{equation}

This formulation preserves partial overlap between alternative derivations. However, each witness of one query may need to be compared with every witness of the other query, resulting in a worst-case number of witness comparisons of

\[
\mathcal{O}\!\left(
|\mathrm{Why}(Q_1,I)|
\times
|\mathrm{Why}(Q_2,I)|
\right).
\]


This cost becomes computationally impractical for large witness collections. Thus, we consider two alternatives: Atomic Why-provenance, treating complete witnesses as indivisible, and Blocked Why-provenance similarity, preserving partial witness overlap while restricting comparisons.

\medskip
\noindent\textbf{Atomic Why-provenance.} Each witness \(w\) is treated as an atomic object and assigned a deterministic identifier derived from its constituent provenance tokens.

\begin{equation}
\mathrm{Why}_{\mathrm{atomic}}(Q,I)
=
\left\{
\mathsf{hash}(w)
\mid
w \in \mathrm{Why}(Q,I)
\right\}.
\end{equation}

Two query executions can then be compared using the Jaccard similarity over their sets of witness identifiers:

\begin{equation}
    \mathrm{sim}_{\mathsf{AW}}(Q_1,Q_2)
=
\frac{
\left|
\mathrm{Why}_{\mathsf{atomic}}(Q_1,I)
\cap
\mathrm{Why}_{\mathsf{atomic}}(Q_2,I)
\right|
}{
\left|
\mathrm{Why}_{\mathsf{atomic}}(Q_1,I)
\cup
\mathrm{Why}_{\mathsf{atomic}}(Q_2,I)
\right|
}.
\end{equation}


This representation reduces similarity computation costs by replacing pairwise witness evaluations with standard set operations. However, it only recognises identical witnesses. Thus, witnesses sharing only some contributing tuples are treated as distinct, effectively ignoring partial overlaps.

\medskip
\noindent\textbf{Blocked Why-provenance similarity.} This approach employs a two-stage candidate selection to preserve partial witness overlap while reducing comparisons.

At the query level, the evaluated execution's Lineage is compared against training execution Lineages. The \(R\) most similar executions are retained as initial candidates. These are subsequently filtered using inexpensive structural properties of their witness collections, including average cardinality and size distributions. At most \(C_q\) candidates proceed to witness-level comparison. If none satisfy these filters, the highest-ranked Lineage candidates are retained as a fallback to ensure a prediction.

At the witness level, comparisons are restricted to witnesses sharing at least one provenance token. For each source witness \(w\), at most \(C_w\) candidates are examined. Let \(\mathcal{C}_{C_w}(w,Q_2)\) denote the witnesses selected from execution \(Q_2\) as candidates for \(w\). The directed blocked similarity between \(Q_1\) and \(Q_2\) is defined as:




\begin{equation}
\mathrm{sim}^{\rightarrow}_{\mathrm{BW},C_w}(Q_1,Q_2)
=
\frac{1}{|\mathrm{Why}(Q_1,I)|}
\sum_{w \in \mathrm{Why}(Q_1,I)}
\max_{v \in \mathcal{C}_{C_w}(w,Q_2)}
\frac{|w \cap v|}{|w \cup v|}.
\label{eq:directed-blocked-why}
\end{equation}

When no candidate witness is available for \(w\), its contribution to the sum is defined as zero. Since the directed measure is not necessarily symmetric, the final blocked similarity is computed as

\begin{equation}
    \mathrm{sim}_{\mathrm{BW}}(Q_1,Q_2)
=
\frac{
\mathrm{sim}^{\rightarrow}_{\mathrm{BW},C_w}(Q_1,Q_2)
+
\mathrm{sim}^{\rightarrow}_{\mathrm{BW},C_w}(Q_2,Q_1)
}{2}.
\end{equation}

This procedure reduces the number of pairwise witness comparisons while retaining sensitivity to partial overlap between derivations. Atomic
Why-provenance defines a standalone representation and can therefore be used both for direct similarity calculation and as a feature representation for machine learning models. Blocked Why-provenance instead defines a pair-dependent comparison procedure and is consequently used only by similarity-based nearest-neighbour methods.

\subsection{Vector-Based Provenance Representations}
\label{subsec:vector-representations}

Query-level provenance abstractions can also be encoded as numerical vectors for conventional classification. Set membership uses binary coordinates, while bag-based provenance preserves token and witness frequencies. Blocked Why-provenance is not vectorised because its comparison requires dynamically matching witnesses across query executions.


\subsubsection{Binary Set-Based Features}

The set-based Lineage and Atomic Why-provenance representations can be encoded as sparse binary vectors. Each coordinate represents a provenance element, and its value indicates whether that element occurs in the corresponding query execution.

\noindent\textbf{Binary Lineage.} Let
%
$\mathcal{U}_{\mathsf{Lin}}
=
\{u_1,\ldots,u_m\}
$
%
denote a common feature space of Lineage tokens. A query execution \(Q\) is encoded as a binary vector
%
$    \boldsymbol{\phi}^{\mathsf{Lin}}(Q)
\in
\{0,1\}^{m},
$
%
whose \(i\)-th coordinate is defined as:

\begin{equation}
\phi_i^{\mathsf{Lin}}(Q)
=
\begin{cases}
1, & \text{if } u_i \in \mathrm{Lin}(Q,I),\\
0, & \text{otherwise}.
\end{cases}
\end{equation}

This vector preserves the membership information of the set-based Lineage representation. It records which distinct input tuples contribute to the query result but does not retain their multiplicity.

\noindent\textbf{Binary Atomic Why-provenance.} This representation is encoded analogously by assigning a common coordinate to each witness identifier. Let
%
$
\mathcal{U}_{\mathsf{Atomic}}
=
\{a_1,\ldots,a_r\}
$
%
denote the corresponding feature space. A query execution \(Q\) is represented by \(\boldsymbol{\phi}^{\mathsf{Atomic}}(Q)\in\{0,1\}^{r}\), whose \(j\)-th coordinate is defined as:

\begin{equation}
\phi_j^{\mathsf{Atomic}}(Q)
=
\begin{cases}
1, & \text{if } a_j \in \mathrm{Why}_{\mathsf{atomic}}(Q,I),\\
0, & \text{otherwise}.
\end{cases}
\label{eq:binary-atomic-why}
\end{equation}

This vector records which distinct witnesses occur in the Why-provenance of the query execution, but not how often they occur.

Both binary representations are stored sparsely, with only active coordinates represented explicitly. They preserve the membership
information contained in the corresponding sets while allowing their use as input to feature-based classification models. These definitions
describe the conceptual feature spaces associated with the two set-based representations. Their bounded implementation using feature hashing is presented in Section~\ref{subsec:implementation-details}.

\subsubsection{Bag-Based Provenance Features}
\label{subsec:bag-representations}

While set-based provenance representations capture the structural contribution of tuples and witnesses, they disregard how often these
elements participate in producing the query result. Bag-based provenance instead preserves this multiplicity information, allowing frequency
effects, such as repeated tuple participation and derivation fan-out, to influence query characterisation directly. Building on the bag-level
abstractions introduced earlier, we derive two frequency-based vector representations.

\noindent\textbf{Token Term Frequency (Token TF).}
This representation describes each execution using token frequencies from its bag-based Lineage. Let \(U=\{t_1,\ldots,t_m\}\) denote the feature space of all workload provenance tokens. Each query \(Q\) is encoded as a sparse vector \(\mathbf{x}(Q)\in\mathbb{R}_{\geq 0}^{m}\), with \(i\)-th coordinate


\begin{equation}
    x_i(Q)=c_Q(t_i),
\end{equation}

\noindent
where \(c_Q(t_i)\) is the multiplicity of token \(t_i\) in \(\mathrm{Lin}_{\mathsf{bag}}(Q,I)\). Unlike binary Lineage, Token TF records not just token presence, but participation frequency.


\noindent\textbf{Witness Term Frequency (Witness TF).} This representation treats witnesses as atomic objects described by their result frequencies. Let \(W=\{w_1,\ldots,w_r\}\) denote the feature space of all distinct workload witnesses. Each query \(Q\) is encoded as a sparse vector \(\mathbf{z}(Q)\in\mathbb{R}_{\geq 0}^{r}\), with \(j\)-th coordinate


\begin{equation}
   z_j(Q)=c_Q(w_j), 
\end{equation}

\noindent
where \(c_Q(w_j)\) is the multiplicity of witness \(w_j\) in \(\mathrm{Why}_{\mathsf{bag}}(Q,I)\). Unlike Binary Atomic Why-provenance, Witness TF captures participation frequency rather than just presence.



These vectors can serve directly as machine learning features or be compared via cosine similarity. For feature vectors \(\mathbf{x}\) and \(\mathbf{y}\), cosine similarity is

\begin{equation}
\mathrm{sim}_{\mathsf{cos}}(\mathbf{x},\mathbf{y})
=
\frac{
\mathbf{x}^{\mathsf{T}}\mathbf{y}
}{
\lVert\mathbf{x}\rVert_2
\lVert\mathbf{y}\rVert_2
}.
\end{equation}

Cosine similarity compares vector directions rather than magnitudes. This emphasises relative feature distributions while mitigating uniform differences in occurrence counts. These definitions describe conceptual feature spaces; their bounded implementation using feature hashing and TF--IDF weighting appears in Section~\ref{subsec:implementation-details}.


\subsubsection{Provenance Structural Descriptors}
\label{subsec:psd}

While the preceding representations retain detailed provenance information at the level of individual tokens or witnesses, it is also possible to derive compact descriptors that summarise the overall structure of query provenance. We introduce Provenance Structural Descriptors (PSD), a vector of simple statistics designed to capture the size of the derivation space and the degree of tuple reuse.

Let \(W(Q)\) denote the set of distinct witnesses, \(W_{\mathsf{tot}}(Q)\) the multiset of all witness occurrences, \(T(Q)\) the set of distinct provenance tokens, and \(T_{\mathsf{tot}}(Q)\) the multiset of all token occurrences.

\begin{equation}
\boldsymbol{\phi}^{\mathrm{PSD}}(Q)
=
\left[
|\mathcal{W}(Q)|,
|\mathcal{W}_{\mathrm{tot}}(Q)|,
|\mathcal{T}(Q)|,
|\mathcal{T}_{\mathrm{tot}}(Q)|,
\frac{|\mathcal{T}(Q)|}{|\mathcal{T}_{\mathrm{tot}}(Q)|},
\frac{|\mathcal{W}(Q)|}{|\mathcal{W}_{\mathrm{tot}}(Q)|},
\frac{|\mathcal{W}(Q)|}{|\mathcal{T}(Q)|}
\right].
\label{eq:psd}
\end{equation}

The first four descriptors capture absolute distinct and total counts of witnesses and provenance tokens. The ratios \(\frac{|\mathcal{T}(Q)|}{|\mathcal{T}_{\mathrm{tot}}(Q)|}\) and \(\frac{|\mathcal{W}(Q)|}{|\mathcal{W}_{\mathrm{tot}}(Q)|}\) measure distinct-to-total proportions for tokens and witnesses, where lower values indicate greater repetition. The final ratio, \(\frac{|\mathcal{W}(Q)|}{|\mathcal{T}(Q)|}\), relates distinct derivations to distinct contributing tuples.


Unlike the sparse token- and witness-level representations, PSD provides a standalone summary with a small, fixed dimensionality that does not depend on the number of distinct provenance elements observed in the workload.

\section{Experimental Evaluation}
\label{sec:results}

We evaluate the effectiveness of the proposed provenance-based representations for identifying the query template that generated a query execution. Specifically, given the provenance associated with an execution, the objective is to predict the TPC-DS template from which the corresponding query was instantiated.

We consider two classification strategies: similarity-based nearest-neighbour classification, which compares query executions using representation-specific similarity measures, and supervised classification, which operates on provenance-derived feature vectors. Both strategies are evaluated using the same experimental protocol and data partitions. The code used in our experiments is publicly available on GitHub\footnote{\url{https://github.com/PauloPintor/provenance-query-similarity}}.



\subsection{Experimental Setup}
\label{subsec:experimental-setup}

TPC-DS~\cite{tpcds} provides parameterised decision-support templates with reproducible class labels. Using a database generated at scale factor \(1\), we generated \(1{,}250\) distinct SQL instantiations for each of ten templates, excluding duplicates to obtain \(12{,}500\) queries. This setting tests template recognition under parameter variation, rather than semantic equivalence between arbitrary SQL queries.

Each base tuple received a unique provenance token, and each query was executed once. Provenance polynomials were obtained through query rewriting following DataProv~\cite{PintorIEEE}, then processed into Witnesses and Lineage and aggregated as defined in Section~\ref{sec:representations}. Each classification instance thus represents the provenance of a complete query result.

A fixed, stratified \(80\%/20\%\) split produced \(10{,}000\) training and \(2{,}500\) test executions, balanced across the ten templates and shared by all methods. We evaluated deterministic, nested training subsets of \(500\), \(1{,}000\), \(2{,}500\), \(5{,}000\), \(7{,}500\), and \(10{,}000\) executions. Each subset was balanced and contained all executions from the preceding smaller subset; the test set remained unchanged. Experiments ran on a 12th-generation Intel Core i7-1265U with \(32\,\mathrm{GB}\) RAM under Arch Linux.

\subsection{Implementation Details}
\label{subsec:implementation-details}

Experiments were implemented in Python. Each Parquet record stored one execution's template label and materialised provenance; only fields needed by the selected representation were loaded. Configurations ran in separate processes to isolate retained state and memory, with results stored after completion.

Nearest-neighbour classification used \(k=1\) and \(k=5\), with unweighted majority voting. Ties were resolved in favour of the class represented by the closest neighbour among the tied classes. Direct Lineage and Atomic Why-provenance retained their complete set-based representations and used Jaccard similarity. Blocked Why-provenance retained complete witness collections and followed the specialised comparison procedure in Section~\ref{sec:representations}. Vector-based nearest-neighbour classification used cosine distance.

The vector representations were also evaluated using Logistic Regression, Multinomial Naive Bayes, Random Forest, and XGBoost. Scikit-learn provided the classifiers and vector-based nearest-neighbour search, except for XGBoost. The supervised classifiers retained their library-default hyperparameters in the main experiments, except for the random seed where applicable; no hyperparameter search was performed. Seed \(13\) controlled training-subset selection and model initialisation.

Feature hashing bounded vector dimensionality without maintaining explicit feature dictionaries. A training-only sensitivity analysis selected the smallest tested power-of-two dimension yielding a mean within-execution collision rate of approximately \(3\%\) or less. This rate was defined as one minus the ratio of occupied coordinates to distinct original features. Binary Lineage and Token TF used \(2^{15}=32{,}768\) coordinates; Binary Atomic Why-provenance and Witness TF used \(2^{14}=16{,}384\). Dimensions remained fixed across training sizes and classifiers.

Token TF and Witness TF received TF--IDF weighting. For vector-based \(k\)-NN and Logistic Regression, sparse features were scaled without mean centring, while dense PSD features were centred and scaled. IDF weights and standardisation parameters were fitted exclusively on training data and applied unchanged to test data.

Blocked Why-provenance used \(R=50\) initial Lineage candidates and at most \(C_q=25\) query candidates after filtering, with a relative tolerance of \(0.75\) on average witness cardinality and a minimum cosine similarity of \(0.90\) between witness-size histograms. If all candidates were rejected, the highest-ranked Lineage candidates were retained. At most \(C_w=25\) witnesses sharing tokens with each source witness were examined.

\subsection{Experimental Results}
\label{subsec:experimental-results}

We report accuracy (\textit{Acc}), macro-averaged F1 (\textit{F1}), which weights templates equally, and mean amortised test-processing time per execution:
\begin{equation}
t_{\mathrm{test/execution}}
=
\frac{t_{\mathrm{build,test}} + t_{\mathrm{predict}}}{N_{\mathrm{test}}}.
\label{eq:test-processing-time}
\end{equation}

\noindent
Here, \(t_{\mathrm{build,test}}\) and \(t_{\mathrm{predict}}\) are the times to construct and transform the test representations and classify the test set, respectively, both expressed in milliseconds; \(N_{\mathrm{test}}\) is the number of test executions.

Vector-based timings include test-feature construction, training-fitted transformations, and prediction. Direct methods operate on in-memory provenance and include similarity calculation and voting. Parquet loading, training-representation construction, preprocessing fitted on training data, and model fitting are excluded. These are amortised test-processing costs, not isolated online latencies or end-to-end runtimes.

Training and test instances are disjoint. Test executions were not used to select training instances or hash dimensions, estimate transformations, or fit classifiers.

\subsubsection{Overall Classification Effectiveness and Processing Cost}
\label{subsubsec:overall-classification-results}



Table~\ref{tab:effectiveness} reports results with \(10{,}000\) training and \(2{,}500\) test executions, with bold identifying the best value within each metric row. For the nearest-neighbour configurations, Lineage and Atomic Why-provenance are compared directly in their set-based form, whereas Token TF, Witness TF, and PSD use vector representations. For the remaining classifiers, Lineage and Atomic Why-provenance use binary feature-hashed encodings, Token TF and Witness TF use TF--IDF-weighted vectors, and PSD is represented as a dense numerical vector. Blocked Why-provenance is evaluated exclusively through nearest-neighbour classification.

\begin{table}[h]
\centering
\caption{Classification effectiveness and mean test-processing time with \(10{,}000\) training and \(2{,}500\) test executions.}
\label{tab:effectiveness}

\begin{tabular*}{\columnwidth}{@{\extracolsep{\fill}}l l c c c c c c}
\toprule
Classifier & Metric & PSD & Lineage &
\shortstack{Atomic \\ Why} &
\shortstack{Token \\ TF} &
\shortstack{Witness \\ TF} &
\shortstack{Blocked \\ Why} \\
\midrule

\multirow{3}{*}{1-NN}
 & Acc & 0.9928 & \textbf{0.9968} & 0.9672 & 0.9272 & 0.8796 & 0.9958 \\
 & F1 & 0.9928 & \textbf{0.9968} & 0.9671 & 0.9175 & 0.8623 & 0.9958 \\
 & \shortstack{Mean test time \\ (ms/execution)} & \textbf{0.0853} & 283.6 & 192.6 & 7.238 & 5.031 & 4315.7 \\
\midrule

\multirow{3}{*}{5-NN}
 & Acc & 0.9836 & \textbf{0.9972} & 0.9452 & 0.8908 & 0.8000 & 0.9968 \\
 & F1 & 0.9836 & \textbf{0.9972} & 0.9462 & 0.8706 & 0.7686 & 0.9968 \\
 & \shortstack{Mean test time \\ (ms/execution)} & \textbf{0.0969} & 283.2 & 192.6 & 7.123 & 5.008 & 5949.4 \\
\midrule

\multirow{3}{*}{LogReg}
 & Acc & 0.9148 & \textbf{0.9996} & 0.9068 & 0.9636 & 0.8848 & -- \\
 & F1 & 0.9132 & \textbf{0.9996} & 0.9023 & 0.9637 & 0.8761 & -- \\
 & \shortstack{Mean test time \\ (ms/execution)} & \textbf{0.0009} & 2.540 & 1.661 & 3.584 & 2.380 & -- \\
\midrule

\multirow{3}{*}{\shortstack{Naive \\ Bayes}}
 & Acc & 0.9540 & 0.9636 & 0.7780 & \textbf{0.9704} & 0.8580 & -- \\
 & F1 & 0.9513 & 0.9639 & 0.7616 & \textbf{0.9701} & 0.8438 & -- \\
 & \shortstack{Mean test time \\ (ms/execution)} & \textbf{0.0007} & 2.420 & 1.687 & 3.586 & 2.391 & -- \\
\midrule

\multirow{3}{*}{\shortstack{Random \\ Forest}}
 & Acc & \textbf{0.9980} & 0.9960 & 0.8872 & 0.9972 & 0.9164 & -- \\
 & F1 & \textbf{0.9980} & 0.9960 & 0.8706 & 0.9972 & 0.9132 & -- \\
 & \shortstack{Mean test time \\ (ms/execution)} & \textbf{0.0045} & 2.904 & 2.019 & 4.086 & 2.710 & -- \\
\midrule

\multirow{3}{*}{XGBoost}
 & Acc & \textbf{0.9988} & 0.9944 & 0.8476 & 0.9948 & 0.9172 & -- \\
 & F1 & \textbf{0.9988} & 0.9944 & 0.8420 & 0.9948 & 0.9168 & -- \\
 & \shortstack{Mean test time \\ (ms/execution)} & \textbf{0.0024} & 2.416 & 1.688 & 3.578 & 2.396 & -- \\
\bottomrule
\end{tabular*}
\end{table}

With \(5\)-NN, Lineage and Blocked Why-provenance achieve accuracy and macro-F1 scores of \(0.9972\) and \(0.9968\). Blocked Why-provenance nevertheless requires approximately \(5949\) ms per execution, compared with \(283\) ms for direct Lineage, making its mean test-processing cost approximately \(21\) times higher.



Increasing the neighbourhood size from one to five raises Lineage accuracy from \(0.9968\) to \(0.9972\), but reduces both accuracy and macro-F1 for PSD, Atomic Why-provenance, Token TF, and Witness TF. Token TF accuracy decreases from \(0.9272\) to \(0.8908\), and Witness TF from \(0.8796\) to \(0.8000\). Therefore, a larger neighbourhood is therefore not consistently beneficial.

Binary Lineage with Logistic Regression achieves \(0.9996\) accuracy and macro-F1, with one misclassified execution and a mean test-processing time of \(2.540\) ms per execution. Compared with direct Lineage using \(5\)-NN, it makes six fewer errors and reduces processing time by a factor of approximately \(111\).


PSD performs best with XGBoost (\(0.9988\) accuracy and macro-F1), followed by Random Forest (\(0.9980\)), at \(0.0024\) and \(0.0045\) ms per execution, respectively. It has the lowest measured test-processing time within every classifier row, although its accuracy varies from \(0.9148\) with Logistic Regression to \(0.9988\) with XGBoost.


Token TF performs best with Random Forest (\(0.9972\) accuracy and macro-F1), matching direct Lineage with \(5\)-NN at \(4.086\) rather than \(283.191\) ms per execution. XGBoost reaches \(0.9948\) for both metrics. Token TF also leads the Multinomial Naive Bayes configurations, with \(0.9704\) accuracy and \(0.9701\) macro-F1.



Atomic Why-provenance performs best with direct \(1\)-NN (\(0.9672\) accuracy, \(0.9671\) macro-F1), whereas Witness TF performs best with XGBoost (\(0.9172\), \(0.9168\)). Both remain below the strongest Lineage, PSD, and Token TF configurations. However, Atomic Why-provenance outperforms Token TF in both nearest-neighbour configurations, while Token TF leads with Logistic Regression, Multinomial Naive Bayes, Random Forest, and XGBoost.

\subsubsection{Effect of Training-History Size}
\label{subsubsec:training-history}

\begin{figure*}[t]
    \centering
    \begin{subfigure}[t]{0.49\textwidth}
        \centering
        \includegraphics[width=\linewidth]{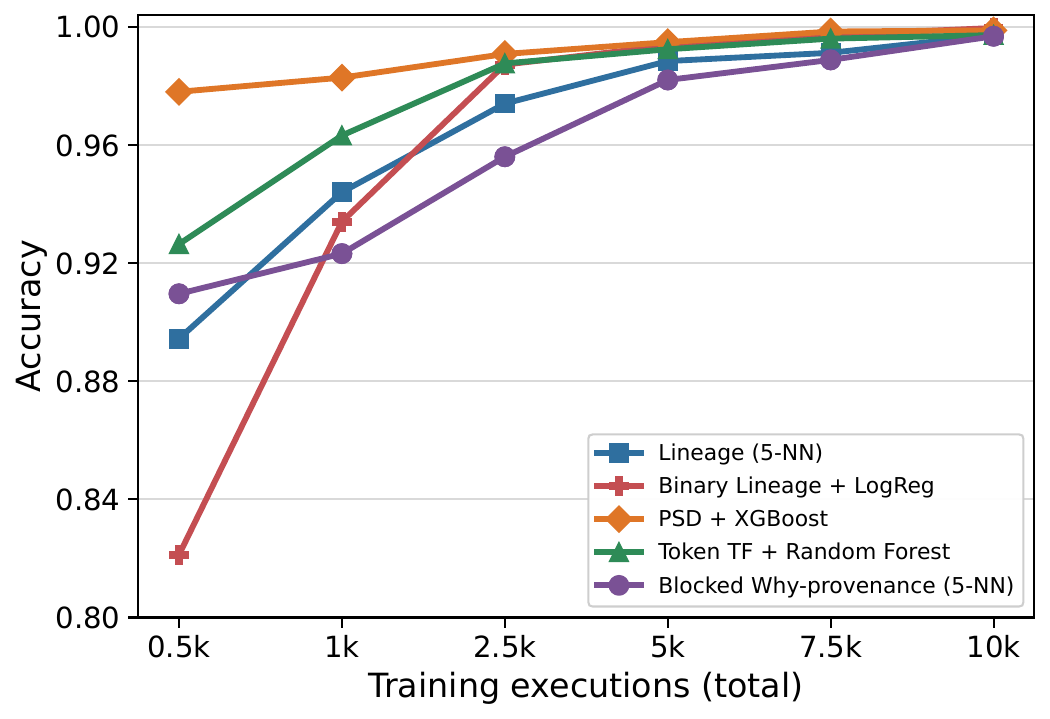}
        \caption{Classification accuracy.}
        \label{fig:training-accuracy}
    \end{subfigure}
    \hfill
    \begin{subfigure}[t]{0.49\textwidth}
        \centering
        \includegraphics[width=\linewidth]{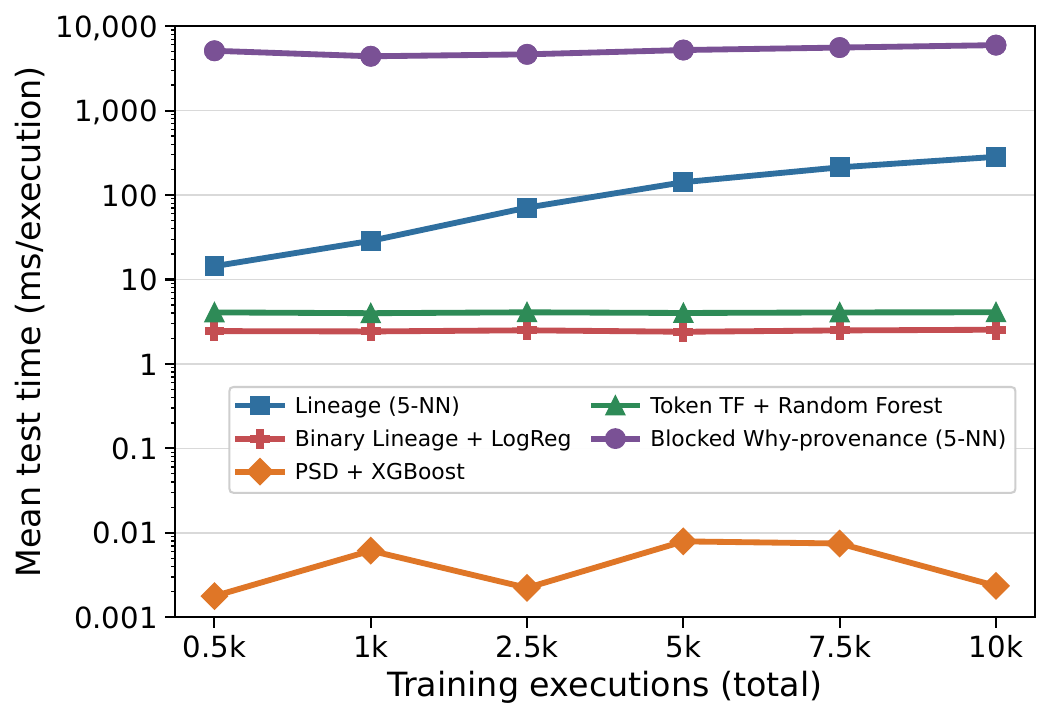}
        \caption{Mean test-processing time.}
        \label{fig:training-time}
    \end{subfigure}
    \caption{Effect of training-history size.}
    \label{fig:training-results}
\end{figure*}

Figure~\ref{fig:training-results} reports accuracy and mean test-processing time for training histories ranging from \(500\) to \(10{,}000\) executions, balanced across the ten templates, with a fixed test set of \(2{,}500\) executions. We compare direct Lineage and Blocked Why-provenance with \(5\)-NN, Binary Lineage with Logistic Regression, PSD with XGBoost, and Token TF with Random Forest.

Figure~\ref{fig:training-accuracy} shows that PSD with XGBoost achieves \(0.9780\) accuracy with only \(500\) training executions. It remains the strongest displayed configuration through \(7{,}500\) executions and reaches \(0.9988\) with the complete history, demonstrating high effectiveness even with limited training data.

The other configurations benefit more substantially from additional history. Binary Lineage with Logistic Regression shows the largest improvement, from \(0.8212\) to \(0.9996\), achieving the highest accuracy at \(10{,}000\) training executions. Direct Lineage, Blocked Why-provenance, and Token TF also improve, narrowing the differences between configurations. With the complete history, all five achieve accuracies between \(0.9968\) and \(0.9996\).

Figure~\ref{fig:training-time} shows contrasting cost trends on a logarithmic scale. Direct Lineage time grows approximately in proportion to training-history size, while Blocked Why-provenance remains the most expensive despite an initial decrease. Binary Lineage and Token TF remain comparatively stable at approximately \(2.5\) and \(4.0\)~ms per execution, respectively, because constructing test representations dominates their classification cost. PSD has the lowest cost throughout, remaining below \(0.008\)~ms per execution.

\subsubsection{Analysis of Provenance Structural Descriptors}
\label{subsubsec:psd-analysis}

We examine the contribution of PSD descriptors through SHAP attributions, descriptor-group comparisons, and single-descriptor experiments. All analyses use \(10{,}000\) training executions and the same fixed test set of \(2{,}500\) executions.

\begin{figure*}[t]
    \centering

    \begin{subfigure}[t]{0.80\textwidth}
        \centering
        \includegraphics[width=\linewidth]{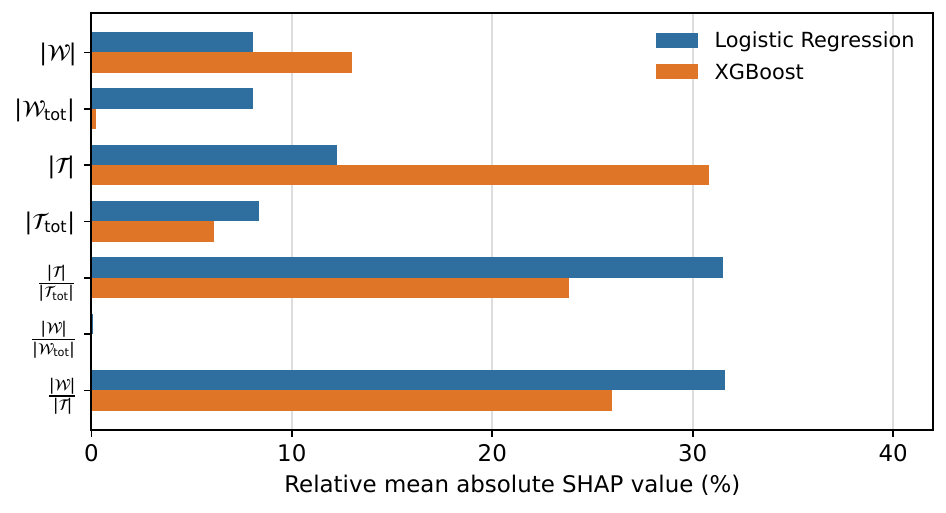}
        \caption{Global importance.}
        \label{fig:psd-shap-global}
    \end{subfigure}

    \par\medskip

    \begin{subfigure}[t]{0.85\textwidth}
        \centering
        \includegraphics[width=\linewidth]{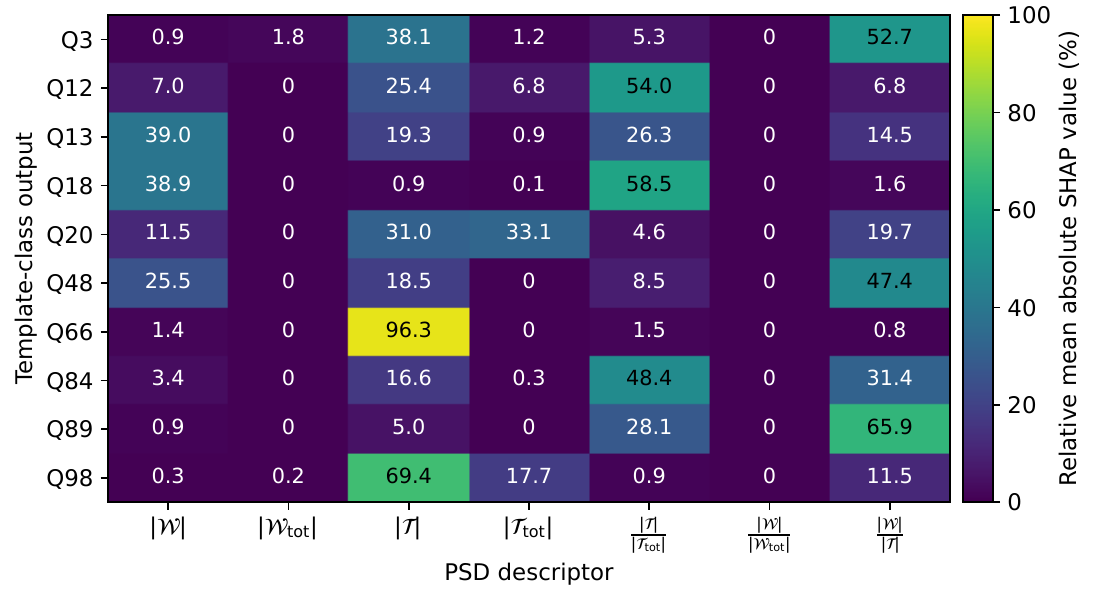}
        \caption{XGBoost importance by template class.}
        \label{fig:psd-shap-by-template}
    \end{subfigure}

    \caption{SHAP importance of PSD descriptors.}
    \label{fig:psd-shap-analysis}
\end{figure*}

Figure~\ref{fig:psd-shap-global} compares global SHAP importance for XGBoost and Logistic Regression. Absolute attributions to class-specific decision scores are averaged across test executions and class outputs, then normalised separately for each model. The percentages describe attribution within each fitted model, not the proportion of classification accuracy attributable to each descriptor.



For XGBoost, distinct token count (\(|\mathcal{T}(Q)|\)) has the highest relative importance at \(30.8\%\). The witness-to-token ratio (\(|\mathcal{W}(Q)|/|\mathcal{T}(Q)|\)) and token-repetition ratio (\(|\mathcal{T}(Q)|/|\mathcal{T}_{\mathrm{tot}}(Q)|\)) follow with \(26.0\%\) and \(23.8\%\), respectively, while distinct witness and total token counts contribute \(13.0\%\) and \(6.1\%\). For Logistic Regression, these ratios lead at \(31.6\%\) and \(31.5\%\), followed by distinct token (\(12.2\%\)), total token (\(8.4\%\)), and distinct witness counts (\(8.1\%\)).

Total witness count (\(|\mathcal{W}_{\mathrm{tot}}(Q)|\)) receives \(0.25\%\) of the attribution in XGBoost but \(8.1\%\) in Logistic Regression. The witness-repetition ratio (\(|\mathcal{W}(Q)|/|\mathcal{W}_{\mathrm{tot}}(Q)|\)) receives zero and \(0.09\%\), respectively. Distinct and total witness counts coincide in \(98.93\%\) of training and \(99.04\%\) of test executions, making this ratio equal to one in those cases. All \(24\) test executions with repeated witnesses belong to Q18.

Figure~\ref{fig:psd-shap-by-template} reports XGBoost attributions by class output, averaged over the full test set and normalised across descriptors. Distinct token count dominates Q66 (\(96.3\%\)) and Q98 (\(69.4\%\)). The witness-to-token ratio leads for Q89 (\(65.9\%\)) and Q3 (\(52.7\%\)), and the token-repetition ratio for Q12 (\(54.0\%\)) and Q18 (\(58.5\%\)). Q48 shows a more distributed pattern: the witness-to-token ratio contributes \(47.4\%\), followed by distinct witness (\(25.5\%\)) and distinct token counts (\(18.5\%\)).

To assess predictive performance beyond attribution, we retrain four classifiers with three variants, named \textit{PSD-Counts}, \textit{PSD-Reduced}, and  \textit{PSD-Full}. 

PSD-Counts contains distinct witness count, total witness count, distinct token count, and total token count. PSD-Reduced contains five descriptors: distinct witness count, distinct and total token counts, the token-repetition ratio, and the witness-to-token ratio. PSD-Full contains all the features used in PSD-Counts and PSD-Reduced.

Preprocessing is refitted on each variant's training features, using the same partitions. For the descriptor-group experiments, the Logistic Regression iteration limit was increased from \(100\) to \(1{,}000\), based solely on training-data convergence checks. All three variants then converged, whereas PSD-Counts and PSD-Full had previously reached the iteration limit. Other settings were unchanged, and Table~\ref{tab:effectiveness} retains the default-setting results.


\begin{table}[t]
    \centering
    \caption{Macro-F1 for the three PSD variants.}
    \label{tab:psd-feature-groups}
    \small
    \begin{tabular}{lccc}
        \toprule
        Classifier & PSD-Counts & PSD-Full & PSD-Reduced \\
        \midrule
        XGBoost             & 0.9944 & 0.9988 & 0.9988 \\
        Random Forest       & 0.9976 & 0.9980 & 0.9980 \\
        \(5\)-NN            & 0.9932 & 0.9836 & 0.9844 \\
        Logistic Regression & 0.8311 & 0.9139 & 0.9139 \\
        \bottomrule
    \end{tabular}
\end{table}

Table~\ref{tab:psd-feature-groups} shows that, compared with PSD-Counts, PSD-Full improves macro-F1 for XGBoost and Logistic Regression, changes Random Forest only slightly, and reduces \(5\)-NN from \(0.9932\) to \(0.9836\). PSD-Reduced preserves PSD-Full's reported macro-F1 for XGBoost, Random Forest, and Logistic Regression; for XGBoost and Logistic Regression, the test predictions are also identical. Its \(5\)-NN score improves slightly to \(0.9844\), remaining below PSD-Counts.

Finally, we conduct an additional experiment with seven XGBoost classifiers, each trained on a single PSD descriptor. None reproduces PSD-Full's \(0.9988\) accuracy. Total witness count performs best alone (\(0.9244\)), while distinct token count reaches \(0.9188\). For Q48, the witness-to-token ratio alone yields a per-template F1 of \(0.9743\). These experiments distinguish a descriptor's importance within the complete model from its predictive performance when used alone.

\subsection{Discussion}
\label{subsec:discussion}

Table~\ref{tab:effectiveness} shows that classification effectiveness depends on the representation and classifier, rather than derivational detail alone. Binary Lineage with Logistic Regression and PSD with XGBoost achieve similarly high effectiveness using tuple identities and aggregate structural properties, respectively. Figure~\ref{fig:training-accuracy} shows that this balance also depends on training-history size. PSD performs strongly with limited data, consistent with learning template-specific structural patterns without relying on token identities, whereas Binary Lineage benefits more from additional examples and achieves the highest accuracy with the complete history.

In particular, greater derivational detail does not necessarily improve classification. The weaker results of Atomic Why-provenance and Witness TF in Table~\ref{tab:effectiveness} are consistent with their treatment of complete witnesses as indivisible features. Unlike Lineage and Token TF, they do not expose shared tokens across different witnesses, and preserving witness frequencies does not restore this information. Blocked Why-provenance recognises partial overlap, but with \(5\)-NN it does not outperform direct Lineage at the largest training size and requires approximately \(21\) times as much test-processing time. Blocking limits comparisons without bounding witness collection sizes, leaving detailed comparison costly.

This computational trade-off also depends on how each representation is used. Figure~\ref{fig:training-time} shows increasing cost for direct Lineage as more historical queries are compared, while the vector-based configurations remain comparatively stable. Nevertheless, fixed dimensionality does not imply fixed prediction cost: different training subsets can produce different tree structures and decision paths in XGBoost. Model differences, witness-candidate selection, and timing variability may contribute to the observed fluctuations, but their effects were not isolated.

The influence of the classifier is also apparent within PSD itself. Table~\ref{tab:psd-feature-groups} shows that adding the ratios benefits Logistic Regression and XGBoost, whereas cosine-based \(5\)-NN performs best with PSD-Counts. Although the ratios add no underlying information to the four counts, they expose their relationships differently. For Logistic Regression, they allow a linear model in the expanded feature space to express nonlinear relationships between counts. For \(5\)-NN, they alter the geometry used to select neighbours, even after standardisation. Their inclusion should therefore be assessed with the classifier rather than assumed universally beneficial.

This does not mean that every descriptor is needed even when the ratios improve classification. PSD-Reduced preserves the reported macro-F1 of XGBoost, Random Forest, and Logistic Regression in Table~\ref{tab:psd-feature-groups} after jointly removing total witness count and the witness-repetition ratio. The near-equality of distinct and total witness counts reported in Section~\ref{subsubsec:psd-analysis} helps explain this redundancy. However, low importance within a complete model does not imply poor standalone performance: total witness count receives little SHAP attribution in XGBoost (Figure~\ref{fig:psd-shap-global}), yet performs best in the single-descriptor experiments. No individual descriptor matches PSD-Full. These findings support selecting combinations of descriptors for the classifier and workload, without establishing that five descriptors are minimal or that witness repetition is uninformative elsewhere.

The scope of these findings ultimately depends on the distinctions required by the task. Different contributing tuples can produce identical PSD values, while identical Lineage can conceal different witness combinations, as illustrated in Section~\ref{subsec:provenance-comparison-motivation}. Aggregate properties distinguish the evaluated templates effectively, but this does not establish a universal preference for summaries over identity-based representations. Tasks requiring distinctions between contributing tuples or their combinations may need the information retained by Lineage or witness-based representations.

\section{Conclusions and Future Work}
\label{sec:conclusion}

This work investigated query-level provenance for query-template identification through set-based and bag-based aggregation of tuple-level Lineage and Why-provenance. The resulting representations capture contributing tuple identities, witness structure, occurrence frequencies, and aggregate properties, providing a perspective grounded in observed data dependencies rather than query expressions or result values alone. Query-level aggregation makes this information available for classification without requiring correspondences between individual output tuples.

Binary Lineage with Logistic Regression achieved the highest accuracy (\(99.96\%\)), while direct Lineage and Token TF also performed strongly. Atomic Why-provenance and Witness TF were less effective, consistent with their reliance on complete witness identities rather than partial overlap. Blocked Why-provenance recognised such overlap but did not outperform direct Lineage with the complete training history and incurred substantially greater test-processing cost. These findings show that additional derivational detail does not necessarily improve classification.

PSD with XGBoost achieved \(97.8\%\) accuracy with \(500\) training executions and \(99.88\%\) with \(10{,}000\), with low test-processing cost once the summaries were materialised. No individual descriptor matched the complete XGBoost model, while PSD-Reduced preserved the reported macro-F1 of XGBoost, Random Forest, and Logistic Regression with five descriptors. The benefits of derived ratios nevertheless depended on the classifier. These results do not establish a universal preference for PSD: queries involving different tuples or witness combinations can have identical descriptor values. Its favourable balance must therefore be reassessed in other workloads, particularly where classification requires distinctions retained by identity-based representations.

Future work will extend the evaluation beyond the ten TPC-DS templates to additional query families, database scales, and operational workloads, including evolving data and query behaviour. We will investigate classifier-aware descriptor selection, query-level provenance summarisation, and adaptive blocking strategies. Workload clustering, comparative debugging, behavioural profiling, and anomaly detection provide further settings for assessing the value of tuple identity, witness structure, and aggregate statistics.

\bmhead{Acknowledgements}

This Research Unit is funded under project UIDB/00127/2025, and this work is funded within the scope of research unit UID/00326 - Centre for Informatics and Systems of the University of Coimbra (https://doi.org/10.54499/UID/00326/2025), both through national funds provided by FCT - Foundation for Science and Technology, I.P.


\bibliography{sn-bibliography}

\end{document}